\documentclass[prb,twocolumn,floatfix]{revtex4-1}
\usepackage{graphicx}
\usepackage{amsmath, amsthm, amssymb}
\usepackage[caption=false]{subfig}
\usepackage{color}

\begin{document}

\title{Effective Field Theory of Interacting $\pi$-Electrons}

\author{J. D. Barr}
\email{barr@physics.arizona.edu}

\author{C. A. Stafford}
\affiliation{Department of Physics, University of Arizona, 1118 East Fourth Street, Tucson, AZ 85721}

\author{J. P. Bergfield}
\affiliation{Nano-Science Center and Department of Chemistry, University of Copenhagen, Universitetsparken 5, 2100 Copenhagen \O, Denmark}

\begin{abstract}
We develop a $\pi$-electron effective field theory ($\pi$-EFT) wherein the two-body
Hamiltonian for a $\pi$-electron system is expressed in terms of
three effective parameters: the $\pi$-orbital quadrupole moment, the on-site repulsion, and a dielectric constant.
As a first application of this $\pi$-EFT, we develop a model of screening in molecular junctions based on image multipole moments,
and use this to investigate the reduction of the HOMO-LUMO gap of benzene.
Beyond this, we also use $\pi$-EFT to 
% CAS
calculate the differential conductance spectrum of the prototypical benzenedithiol-Au single-molecule junction and the 
$\pi$-electron contribution to the van der Waals interaction between benzene
and a metallic electrode.

\end{abstract}

\maketitle
	
\section{Introduction}
\label{sec:intro}

Owing to the profound versatility of the carbon-carbon bond, organic molecules form the basis for a
myriad of 
potential nanotechnology applications. Many of these
make use of the ability of conjugated organic molecules to conduct electricity, in which case the system of delocalized $\pi$-electrons plays a role analogous to that of the conduction band in a conventional semiconductor. In such devices, the most important degrees of freedom from a technological perspective are
those associated with these current-carrying $\pi$-electrons. 

The main motivation of the present work is to derive a model Hamiltonian for $\pi$-electron systems to facilitate the study of many-body effects on transport 
through molecular heterojunctions.  The standard paradigm for molecular junction transport calculations involves local or semilocal approximations to density functional
theory (DFT) combined with nonequilibrium Green's functions (NEGF).  This DFT-NEGF approach\cite{Cuevas10} has tremendous advantages in terms of computational efficiency
and chemical realism.  However, it has notorious difficulties describing the energetics most relevant for electron transport, % in molecular junctions, 
namely the energy level alignment between molecule and metal electrodes,
and the fundamental (or HOMO-LUMO) gap.
Some possible underlying reasons for this are (i) the failure to include nonlocal correlations responsible
for screening of intramolecular interactions by nearby metal electrodes;\cite{Neaton06,Thygesen09} 
(ii) self-interaction error;\cite{Toher05,Ke07,Cohen08,Toher08} and (iii) omission of the derivative 
discontinuity\cite{Perdew82,Bergfield12} needed to describe the quantization of the molecular charge within the junction.\cite{Datta06}
Self-consistent many-body perturbation theory\cite{Thygesen11} is able to overcome hurdles (i) and (ii), but still leaves (iii) as an open problem.
%the role of the derivative discontinuity 

%Although the standard paradigm for transport calculations in molecular junctions--based on local approximations within density functional theory (DFT)--includes these degrees of freedom, it does so amongst many others. This comes at the expense of an exact treatment of intramolecular many-body interactions, which
%to date has made it challenging to capture important interaction effects such as Coulomb blockade within the context of DFT.\cite{Datta06}

%Intuitively, this shortcoming can be explained by the apparent inability of local-DFT to capture the particle-like character of the electron.
%For example, the spurious repulsive interaction predicted by most DFT treatments of the dissociation of the hydrogen molecular ion is consistent
%not with the presence a single electron with charge $e$, but rather many electrons,
%each with fractional charge;\cite{Cohen08} ultimately, this is equivalent to the presence of an unphysical self-interaction, but when this issue is understood in terms
%of an incomplete treatment of wave-particle duality, it becomes clear that effects depending strongly on the quantization electric charge--like
%Coulomb blockade--are fundamentally difficult to capture within local-DFT.\cite{Datta06} Indeed, since the central quantity in DFT
%is the overall electron density $n(\vec x)$ rather than a many-body wavefunction $\Psi(\vec{x}_1, \vec{x}_2, \ldots)$ depending upon the coordinates of \emph{all} the electrons, devising
%functionals that account satisfactorily for charge quantization is challenging.

An alternative approach is to formulate a model including only the degrees of freedom essential to describing the $\pi$-electron dynamics,
thereby reducing the overhead associated with %exact diagonalization of the molecular Hamiltonian. 
an exact treatment of interactions within the junction. 
Electron transport can then be treated using many-body Green's function techniques,\cite{Stafford09,Rincon09}
the Master equation approach,\cite{Koenig97,Schoeller00,Wacker05,Harbola08,Galperin08}
or quantum impurity solvers.\cite{Bohr06}
This procedure begins with the observation that processes in systems of $\pi$-electrons
take place at characteristic length, energy and time scales
all ultimately dictated by the strength of the $\pi$-electron bond.
Intuitively, one expects that only degrees
of freedom with scales comparable to these need to be explicitly included. Semi-empirical models based on this notion have been in use for over fifty years,\cite{Pariser53, Parr60, Ohno64, Pople53}
and work to improve their accuracy is ongoing.\cite{Chandross97,Bursill98, Castleton02} However, since these are based on ad hoc
parameterizations\cite{Ohno64, Chandross97,Castleton02, Bursill98, Ramasesha91} of interparticle Coulomb interactions that do not satisfy Maxwell's equations,
it is difficult to extend such techniques to include effects like the screening of intramolecular interactions by the electrodes in molecular junctions.

% MODIFIED: Based on comments 1 and 2 from referee 2
%To avoid the foregoing difficulties one may instead formulate a model of conjugated organic molecules including only the degrees of freedom essential to describing the $\pi$-electron dynamics, permitting an exact treatment of electron correlation and thereby capturing important transport phenomena presently beyond the scope of local-DFT like Coulomb blockade. This procedure begins with the observation that processes in systems of $\pi$-electrons take place at characteristic length, energy and time scales
%all ultimately dictated by the strength of the $\pi$-electron bond. 
%Intuitively, one expects that only degrees
%of freedom with scales comparable to these need to be explicitly included. Semi-empirical models based on this notion have been in use for over fifty years,\cite{Pariser53, Parr60, Ohno64, Pople53}
%and work to improve their accuracy is ongoing.\cite{Bursill98, Castleton02} However, since these are based on ad hoc
%parameterizations\cite{Ohno64, Castleton02, Bursill98, Ramasesha91} of interparticle Coulomb interactions that do not satisfy Maxwell's equations,
%it is difficult to extend such techniques to include effects like the screening of intramolecular interactions by the electrodes in molecular junctions.

\begin{figure}[t]
\includegraphics[scale=0.43]{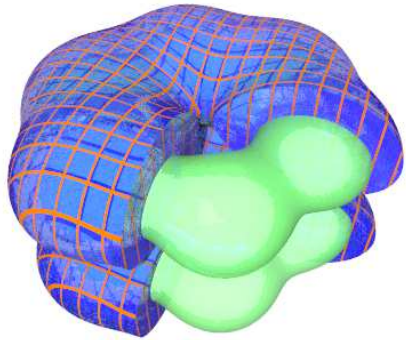}
\caption{\label{fig:isosurfaces} Two isosurfaces of the average $\pi$-electron density $\langle \psi^\dagger(\vec x) \psi(\vec x) \rangle$ depict the electronic structure of gas-phase benzene
within $\pi$-EFT.}
\end{figure}

In contrast to this, effective field theory (EFT) provides a concise, systematic method of constructing a $\pi$-electron 
Hamiltonian starting from first principles by performing an expansion in a small parameter and then
imposing symmetry constraints. The result contains a few physically meaningful parameters, which are then renormalized to include the aggregate effect of the degrees of freedom not explicitly retained. In this article we proceed along
these lines, first expanding the full electronic Hamiltonian of a conjugated organic molecule in a basis of atomic orbitals
and then dropping terms involving energies far from the $\pi$-electron bond energy.

Imposing symmetry constraints and performing an expansion in powers of the interatomic bond length then allows us to construct an effective Hamiltonian for the $\pi$-electrons in a conjugated organic molecule
that accounts for the effects of $\sigma$-electrons virtually.\footnote{Within EFT, the $\sigma$-electrons can be included explicitly at the expense of a larger basis, and this may be necessary in order to explain transport in some experiments.\cite{Venkataraman06}} 
\nocite{Venkataraman06}
As an example of this, we consider the particular case of gas-phase benzene, for which we formulate an effective Hamiltonian with only four adjustable
parameters: the on-site repulsion $U$, the nearest-neighbor hopping matrix element $t$, a dielectric constant $\epsilon$, and the
$\pi$-electron quadrupole moment $Q$. In principle, these could then be renormalized ab initio,
e.g.\ by using perturbation theory to freeze out degrees of freedom far from the $\pi$-electron
energy scale; however, since this is tedious and would not enhance the predictive power of our model,
we instead fit the parameters directly to experiment.

Next, we show how screening from metallic electrodes can be incorporated into this scheme without
introducing additional parameters by considering the multipole moments of image charge distributions.
We then use this method of screening to calculate the screened HOMO-LUMO gap of benzene near a metallic electrode, as well as to formulate a realistic 
model of a gold-benzenedithiol-gold junction, including effects arising from the presence of the thiol sidegroups. 
% CAS
The differential conductance spectrum of the junction is calculated as a function of the gate and bias voltages in the experimentally relevant regime,\cite{Song09}
exhibiting the characteristic diamond-shaped features\cite{Stafford09} indicative of quantized charge on the molecule within the junction.
Finally, we also use this $\pi$-electron effective
field theory ($\pi$-EFT) to compute the $\pi$-electron contribution to the van der Waals interaction between benzene and a metallic electrode.

\section{Bare Hamiltonian}

Using the Born-Oppenheimer approximation, the one-body term in the electronic Hamiltonian for an isolated molecule can be written as
\begin{equation}
\label{eq:hamiltonianField}
H^{(1)} = \sum_{\sigma} \int d^3x \, \psi^\dagger_\sigma(\vec x) \left( \frac{-\hbar^2}{2m} \nabla^2 + V \right) \psi_{\sigma}(\vec x)
\end{equation}
where $V$ is the interaction between the electrons and the atomic nuclei. The operator that creates an electron with spin $\sigma$ in
the $n$th element of a basis of atomic orbitals $\left\{\phi_n\right\}$ can be expressed as:
\[
d^\dagger_{n\sigma}  =  \int d^3x \, \phi_n(\vec x) \psi^\dagger_\sigma(\vec x)
\]
Multiplying this by the inverse of the overlap matrix $S_{nm} = \langle \phi_n | \phi_m \rangle$ and summing over $m$ implies
\begin{align}
\sum_m d^\dagger_{n\sigma} S^{-1}_{nm} \phi^*_m(\vec x) & = \int d^3x' \sum_m \phi_n(\vec x') S^{-1}_{nm} \phi^*_m(\vec x) \psi^\dagger_\sigma(\vec x') \nonumber \\
& = \psi^\dagger_\sigma(x) \label{eq:psiDagger}
\end{align}
where we have made use of the completeness relation for a non-orthogonal basis:\cite{Thygesen06}
\[
\sum_{nm} \phi_n(\vec x') S^{-1}_{nm} \phi_m^*(\vec x) = \delta(\vec x - \vec x')
\]
Combining equations \eqref{eq:hamiltonianField} and \eqref{eq:psiDagger} then gives:
\begin{equation}
\label{oneBodyHamiltonianExact}
H^{(1)} = \sum_{nm\sigma}  \mathcal{H}_{nm} d^\dagger_{n\sigma} d_{m\sigma}
\end{equation}
where
\begin{equation}
\label{generalOnebodyHamiltonian}
H^{(1)}_{nm} = \int d^3x \, \phi^*_n(\vec x) \left( \frac{-\hbar^2}{2m} \nabla^2 + V \right) \phi_m(\vec x)
\end{equation}
and
\begin{equation}
\label{oneBodyMatrixElementsExact}
\mathcal{H}_{nm} = \sum_{kl} S^{-1}_{nk} H^{(1)}_{kl} (S^{-1}_{ml})^*.
\end{equation}
% MODIFIED: Based on comment from referee 2
If we keep only nearest-neighbor terms this reduces to the H\"uckel Hamiltonian
\[
H^{(1)} = \sum_n \varepsilon_n d^\dagger_{n\sigma} d_{n\sigma} - \sum_{\langle n, m\rangle, \sigma} t_{nm} d^\dagger_{n\sigma} d_{m\sigma}
\]
where $t_{nm} = \mathcal{H}^{(1)}_{nm}$ and $\varepsilon_n = \mathcal{H}^{(1)}_{nn}$.

Similarly, the two-body term in the electronic Hamiltonian can be written as
\begin{align*}
H^{(2)} & =  \frac{1}{2} \sum_{\sigma\sigma'} \int d^3x_1 d^3x_2 \, \psi_{\sigma}^\dagger(\vec x_1) \psi_{\sigma'}^\dagger(\vec x_2) \frac{e^2}{|\vec x_1 - \vec x_2|} \times \\
& \psi_{\sigma'}(\vec x_2)\psi_{\sigma}(\vec x_1),
\end{align*}
which, in terms of the atomic orbital basis, is equivalent to
% MODIFIED: Based on comment from referee 2
\begin{equation}
\label{twoBodyHamiltonianExact}
H^{(2)} = \frac{1}{2} \sum_{nmlk\sigma \sigma'}  \mathcal{U}_{nmlk}  d^\dagger_{n\sigma} d^\dagger_{m\sigma'} d_{l\sigma'} d_{k\sigma} 
\end{equation}
where
\begin{align*}
U_{nmkl} & = \int d^3x_1 d^3x_2 \,\phi^*_n(\vec x_1)\phi^*_m(\vec x_2) \frac{e^2}{|\vec x_1 - \vec x_2|} \times \\
& \phi_k(\vec x_2) \phi_l(\vec x_1)
\end{align*}
and
\begin{equation}
\label{twoBodyMatrixElementsExact}
\mathcal{U}_{nmkl} = \sum_{opqr} S^{-1}_{no} S^{-1}_{mp}  U_{opqr} (S^{-1}_{kq})^* (S^{-1}_{lr})^*.
\end{equation}

Together, equations \eqref{oneBodyHamiltonianExact}-\eqref{oneBodyMatrixElementsExact} and \eqref{twoBodyHamiltonianExact}-\eqref{twoBodyMatrixElementsExact}
give the full electronic Hamiltonian from first principles:
\[
H = H^{(1)} + H^{(2)},
\]
but do so in terms of a basis that is impractically large for use within existing many-body techniques.  To overcome this difficulty, 
in the next section we formulate an effective Hamiltonian in a reduced basis, explicitly retaining only the degrees of freedom necessary
to describe the $\pi$-electron dynamics. 

\section{Effective Hamiltonian}
\label{sec:eff_ham}
	
The first step in constructing the effective Hamiltonian is culling elements of the basis that lie far from the energy scale of interest. To this
end, we first exclude atomic orbitals that do not participate in chemical bonding (those corresponding to core or excited electrons), which, in a
$\pi$-electron system, leaves an effective $s$ orbital and three effective $p$ orbitals at each atom. The former hybridize with the effective
$p_x$ and $p_y$ orbitals giving rise to three $sp^2$ hybrids that form the $\sigma$ bonds between the atoms. The remaining effective $p_z$ orbitals,
which, for a planar molecule, cannot hybridize with any of the $\sigma$-electrons without breaking inversion symmetry, are occupied by one
electron on each atom and form $\pi$ bonds with weaker binding energies. Because of this energy difference we also omit the atomic orbitals participating
in the $\sigma$ bonds, though this approximation could be relaxed at the expense of a larger basis.

The effective Hamiltonian for the remaining effective $p$ orbitals can then be determined using equations \eqref{oneBodyHamiltonianExact} through \eqref{twoBodyMatrixElementsExact} if the effective orbitals are known. In principle, these could be calculated directly, e.g.\ by using
perturbation theory to freeze out the degrees of freedom far from the $\pi$-electron energy scale; however, as noted previously we find it more
practical to parametrize these expressions by imposing symmetry constraints.

% MODIFIED: Bsaed on comment 3 from referee 2
To do this, we work initially in the asymptotic limit where the interatomic bond length is large compared to the size of the effective orbitals. This
condition implies that matrix elements $U_{nmkl}$ with $n \neq l$ or $m \neq k$ and overlap integrals $S_{nm}$ with $n \neq m$ are exponentially
small, allowing us to reduce the interaction matrix (Eq.\ \eqref{twoBodyMatrixElementsExact}) to
\begin{align}
\label{ndoApprox}
\mathcal{U}_{nmkl} & =  U_{nmkl} \nonumber \\
& =  \delta_{nl} \delta_{mk} \int d^3x_1 d^3x_2 \, \frac{e^2 |\phi_n(\vec x_1)|^2 |\phi_m(\vec x_2)|^2}{|\vec x_1 - \vec x_2|} \nonumber \\
& \equiv  \delta_{nl} \delta_{mk} U_{nm}
\end{align}
where $\phi_n$ are now effective instead of bare orbitals. 
Although it is known\cite{Ohno64} that the terms neglected are not a priori negligible
at typical interatomic distances, it has been suggested that this approximation can be justified by the use of orthogonalized orbitals,\cite{Roby71}
% CAS
and it has been explicitly shown\cite{Baeriswyl92} that this is an accurate approximation
for $\pi$-conjugated systems.
Here we offer a simpler perspective more consistent with the spirit of EFT, namely that the neglected terms are accounted
for virtually when the parameters in the Hamiltonian are renormalized. We also note that Eq.\ \eqref{ndoApprox} is equivalent to the ``neglect of
differential overlap approximation'' that has already been used extensively elsewhere, but that in the context of EFT it is
simply the requirement that the effective Hamiltonian be local. 
%It has been explicitly shown elsewhere\cite{Baeriswyl92} that this is a satisfactory approximation for $\pi$-conjugated systes; 
However, we note here that in order to extend the present work to the case where multiple orbitals (e.g. both $\sigma$ and $\pi$) are
centered on the same atom, it would be necessary to include the same-site interaction matrix elements as additional parameters.

Expanding Eq.\ \eqref{ndoApprox} in powers of the interatomic bond length yields a standard electrostatic multipole expansion, and, if we assume
the effective $p$ orbitals possess azimuthal and inversion symmetry, $U_{nm}$ can be parametrized up to the quadrupole-quadrupole interaction
in terms of the on-site repulsion $U_{nn}$ and the $zz$ component of the quadrupole moment $Q_n$ associated with each orbital, as well as
a dielectric constant $\epsilon$ included to account for the polarizability of the $\sigma$ and core electrons.

Explicitly, this gives
\begin{align}
U_{nm} & =  U_{nn}\delta_{nm} \nonumber \\ 
& +  (1 - \delta_{nm})\left(U^{MM}_{nm} + U^{QM}_{nm} + U^{QM}_{mn} + U^{QQ}_{nm}\right) \nonumber \\
& +  O(r^{-6}),\label{parametrization} 
\end{align}
where $U^{MM}$ is the monopole-monopole interaction, $U^{QM}$ is the quadrupole-monopole interaction, and $U^{QQ}$ is the quadrupole-quadrupole
interaction. For two orbitals with arbitrary quadrupole moments $Q^{ij}_n$ and $Q^{kl}_m$ separated by a displacement $\vec r$, the expressions
for these are
\begin{align}
U^{MM}_{nm} & = \frac{e^2}{\epsilon r} \label{monopoleMonopole} \\
U^{QM}_{nm} & = \frac{-e}{2 \epsilon r^3} \sum_{ij} Q_m^{ij} \hat{r}_i\hat{r}_j  \\
U^{QM}_{mn} & = \frac{-e}{2 \epsilon r^3} \sum_{ij}  Q_n^{ij} \hat{r}_i \hat{r}_j  \\
U^{QQ}_{nm} & = \frac{1}{12 \epsilon r^5} \sum_{ijkl}  Q_n^{ij} Q_m^{kl} W_{ijkl}, \label{quadrupoleQuadrupole}
\end{align}
where
\begin{align*}
W_{ijkl}  & =  \delta_{li} \delta_{kj} + \delta_{ki} \delta_{lj} - 5 r^{-2}(r_k \delta_{li} r_j + r_k r_i \delta_{lj} \\ 
& + \delta_{ki} r_j r_l + r_i \delta_{kj} r_l + r_k r_l\delta_{ij}) + 35 r^{-4} r_i r_j r_l r_k
\end{align*}
is a rank-four tensor that characterizes the interaction of two quadrupoles.\cite{Hernandez-Trujillo93} Altogether, this provides an expression for the interaction energy
that is correct up to fifth order in the interatomic distance.

To further reduce the number of free parameters it is convenient to simplify the effective Hamiltonian by requiring it to satisfy particle-hole symmetry.
Although this is not strictly necessary within the context of $\pi$-EFT, the success of Pariser-Parr-Pople type
semi-empirical models--which implicitly assume particle-hole symmetry--suggests that it is a good approximation to do so. Taking this to be the case, Eq.\ \eqref{generalOnebodyHamiltonian} 
then gives for the one-body Hamiltonian: 
\begin{equation*}
H^{(1)}_{nm} = \int d^3x \, \phi^*_n(\vec x) \left( \frac{-\hbar^2}{2m} \nabla^2 + \sum_l V_l(\vec x)\right) \phi_m(\vec x), 
\end{equation*}
where $V_l(\vec x)$ is the effective potential due to the ionic hole at site $l$:
\[
V_l(\vec x) = \int d^3x' \, \frac{-e^2|\phi_l(\vec x')|^2}{\epsilon |\vec x - \vec x'|}.
\]
Using Eq.\ \eqref{ndoApprox} then gives
\[
H^{(1)}_{nm}  =  \delta_{nm}\left(\varepsilon^{(at)}_n - \sum_{l \ne n} U_{nl}\right) + (1 - \delta_{nm})t_{nm},
\]
where we have defined the atomic on-site energy as:
\[
\varepsilon^{(at)}_n = \int d^3x \, \phi^*_n(\vec x)\left(\frac{-\hbar^2}{2m} \nabla^2 + V_n(\vec x)\right) \phi_n(\vec x).
\]
Defining $\rho_n = \sum_\sigma d^\dagger_{n\sigma} d_{n\sigma}$ and rearranging
the two-body term then yields:
\begin{align*}
H^{(1)} + H^{(2)} & =  \sum_n \varepsilon^{(at)}_n\rho_n - \sum_{\langle n,m \rangle, \sigma} t_{nm} d^\dagger_{n\sigma} d_{m\sigma}  \\
& +  \frac{1}{2} \sum_{nm} U_{nm} (\rho_n - 1)(\rho_m - 1) \\
& +  \frac{1}{2} \sum_{n} U_{nn} \rho_n -  \frac{1}{2} \sum_{nm} U_{nm}.
\end{align*}
Finally, adding the mutual repulsion of the ionic cores $\frac{1}{2} \sum_{n \ne m} U_{nm}$ gives the full effective molecular Hamiltonian:
\begin{align*}
H & =  \sum_n \varepsilon^{(at)}_n \rho_n - \sum_{\langle n, m 
\rangle, \sigma} t_{nm} d^\dagger_{n\sigma} d_{m\sigma}\\
& +  \frac{1}{2} \sum_{nm} U_{nm} q_n q_m  \\
& +  \frac{1}{2} \sum_{n} U_{nn} q_n,
\end{align*}
where we have introduced the effective charge operator defined by $q_n = \rho_n - 1$. In conjunction with Eq.\ \eqref{parametrization}, this
expresses the effective Hamiltonian for an arbitrary $\pi$-electron system in terms of the tight-binding matrix $t_{nm}$, the on-site repulsion
$U_{nn}$, a dielectric constant $\epsilon$, and the $\pi$-electron quadrupole moment $Q_n$.

In the remainder of this paper we focus on benzene as a benchmark system, in which case the Hamiltonian reduces to
\begin{equation}
\label{benzeneEffectiveHamiltonian}
H = \mu \sum_n \rho_n - t\sum_{\langle n, m \rangle, \sigma} d^\dagger_{n\sigma} d_{m\sigma} +  \frac{1}{2} \sum_{nm} U_{nm} q_n q_m,
\end{equation}
where $U_{nn} = U$ and $Q^{zz}_n = -Q^{xx}_n/2 = -Q^{yy}_n/2 \equiv Q$ by symmetry. The molecular chemical potential $\mu$ is fixed by the experimental ionization energy\cite{Howell84, Kovac80, Sell78, Kobayoshi78, Schmidt77}
and electron affinity:\cite{Burrow87}
\[
\mu =  \frac{IE - EA}{2} = -4.06 \; \text{eV},
\]
whereas the four other parameters must be renormalized by fitting to experiment, which is the subject of the following section.

\section{Renormalization: Fitting the gas-phase spectrum}

We have renormalized the parameters in our effective Hamiltonian for gas-phase benzene by fitting to experimental values that should
be accurately reproduced within a $\pi$-electron only model. In particular, we have simultaneously optimized the theoretical predictions of 1) the
vertical ionization energy, 2) the vertical electron affinity, and 3) the six lowest singlet and triplet excitations of the neutral molecule.

\begin{table*}

\begin{tabular}{ l|c|c|c|c }
  & Exp.\cite{Howell84, Kovac80, Sell78, Kobayoshi78, Schmidt77, Burrow87, Hiraya91, Doering69, Frueholz79, Koch72} & $\pi$-EFT & PPP                                & PPP \\
  &                                                                                                                 &           & Castleton et. al\cite{Castleton02} & Ohno\cite{Ohno64,Ramasesha91} \\
\hline
\hline
  Ionization Energy (eV) & 9.23  & 9.26 & 9.05 & 9.78 \\
\hline
  Electron Affinity (eV) & -1.12 & -1.14 & -0.93 & -1.67\\
\hline
  Neutral Spectrum (eV)           &  &  &  \\	
  \hspace{1cm} Singlet            &  4.90 & 4.87 & 4.76 & 4.23 \\	
                                  &  6.21 & 6.08 & 6.30 & 5.52 \\	
                                  &  6.93 & 7.59 & 6.93 & 6.81 \\	
                                  &                  & &      &      \\
  \hspace{1cm} Triplet            &  3.93 & 4.10 & 3.99 & 3.52 \\	
                                  &  4.75    & 4.92 & 4.74 & 4.32 \\
                                  &  5.60 & 6.17 & 5.84 & 5.58 \\

\hline
  RMS Relative Error (\%)               & N/A & 4.2 & 6.0 & 19.0
\end{tabular}

\caption{\label{comparisonTable} Experimental data for the vertical ionization energy,\cite{Howell84, Kovac80, Sell78, Kobayoshi78, Schmidt77} vertical electron affinity,\cite{Burrow87}
and optical spectrum,\cite{Hiraya91, Doering69, Frueholz79, Koch72} of gas-phase benzene compared to the predictions of $\pi$-EFT, a recent
PPP-type model,\cite{Castleton02} and the Ohno parametrization.\cite{Ohno64, Ramasesha91} The best fit parameterization of our $\pi$-EFT was
determined to be $t = 2.70$ eV, $U = 9.69$ eV, $Q = -0.65$ $e$\AA$^2$ and $\epsilon = 1.56$.}

\end{table*}

This was done by exactly diagonalizing Eq.\ \eqref{benzeneEffectiveHamiltonian} with the interatomic bond length\cite{Tamagawa76} fixed at
$1.40 \; \textrm{\AA}$. In particular, using the OQNLP algorithm\cite{Zsolt07} for nonlinear global optimization we
minimized the RMS relative error of our predictions for the quantities in the first column of Table \ref{comparisonTable}. The results of this
procedure, which converged to the same solution regardless of initial conditions, are summarized in column two of the same table.
The optimal parametrization for the $\pi$-EFT was found to be  $t = 2.70$ eV, $U = 9.69$ eV, $Q = -0.65$ $e$\AA$^2$ and $\epsilon = 1.56$
with a RMS relative error of $4.2$ percent.

Also appearing in Table \ref{comparisonTable} are the predictions of a recent Pariser-Parr-Pople type semi-empirical model\cite{Castleton02} as well
as those of the original Ohno parametrization.\cite{Ohno64, Ramasesha91} Compared to the recent PPP model, $\pi$-EFT fits the
optical spectrum of gas-phase benzene to a similar degree of accuracy and gives better results for the ionization energy and
electron affinity. Moreover, the parameters common to both models have comparable values, namely those given above for our model
% CAS
and those of the model of Castleton et al.\cite{Castleton02} ($t = 2.64$ eV, $U = 8.9$ eV, and $\epsilon = 1.28$). The $\pi$-EFT on-site repulsion is also in qualitative
agreement with recent RPA-based calculations of the effective Coulomb repulsion in graphene.\cite{Wehling11}

Although our effective quadrupole moment has no direct counterpart in phenomenological models, its value can be compared to the
bare quadrupole moment of a hydrogenic $2p_z$ orbital, which is given by
\begin{equation}
\label{quadrupoleFromEffectiveCharge}
Q_{zz} = -24 e (a_0 / Z)^2,
\end{equation}
where $a_0$ is the Bohr radius and $+Ze$ is the nuclear charge. Using this, we find that our $\pi$-orbital quadrupole moment
corresponds to a hydrogenic $p$ orbital bound by an effective charge of $+3.22e$, or, equivalently, with an effective Bohr radius of $0.16$ \AA.
This is consistent with the expectation that the $sp^2$ orbitals forming the $\sigma$ bonds provide only weak screening of the atomic core,
which has a net charge of +4e. For the purpose of visualization, effective hydrogenic orbitals can also be used to render the average
$\pi$-electron density $\langle \psi^\dagger(\vec x) \psi(\vec x) \rangle$, as shown in Figure \ref{fig:isosurfaces}.

\section{Screening by metallic electrodes: The image multipole method}

In this section, we extend the preceding model to include the effect of screening by metallic electrodes, which for simplicity are modeled
as planar or spherical conductors. In the regime where the characteristic response time of the electrons in the electrode is much shorter
than the timescale of the $\pi$-electron dynamics, this can be done using the method of images via a straightforward
extension of Eq.\ \eqref{parametrization}. This is expected to be the case for conjugated organic molecules in the vicinity of gold electrodes,
in which case the metallic plasma frequency\cite{Linden04} $\omega_p \approx 9 \; \textrm{eV} /\hbar$ is large compared to the frequency scale
of $\pi$ excitations $\omega_{\pi} \approx 2t / \hbar \approx 5 \; \textrm{eV} / \hbar$. The leading order correction to the metallic dielectric function,
given by the GW approximation, then goes as $(\omega_\pi / \omega_p)^2 \approx 0.3$.  Explicit calculations using the GW approach also
suggest that corrections to the image charge method tend to be small for organic molecules adsorbed on a metallic surface.\cite{Neaton06}

In the following subsections, the multipole moments of the image charge distribution generated by an orbital near planar and spherical
conductors are described. To determine the screened interaction matrix, interactions between these and the orbital multipole moments
are included in $U_{nm}$ using equations \eqref{monopoleMonopole} through \eqref{quadrupoleQuadrupole}. Overall, the two-body Hamiltonian
should give the energy required to prepare the molecular charge distribution by bringing each of the electrons in from infinity with
the electrodes maintained at fixed electrostatic potentials. This can be ensured using a number of different counting schemes, but we
take one that ensures a symmetric interaction matrix, namely
\[
\tilde{U}_{nm} = U_{nm} + \delta_{nm}U_{nn}^{(i)} + \frac{1}{2}(1 - \delta_{nm})(U_{nm}^{(i)} + U_{mn}^{(i)}),
\]
where $U_{nm}$ is the unscreened interaction matrix, $U_{nm}^{(i)}$ is the interaction between the $n$th orbital and the image of the $m$th orbital,
and $\tilde{U}_{nm}$ is the screened interaction matrix. Since the image multipole moments of an orbital change as it is
brought in from infinity, one might expect a prefactor of $1/2$ in the second term of the preceding equation, however, this is already
present in the Hamiltonian itself.

When multiple electrodes are present, the image of an orbital in one conductor produces images in the other electrodes, resulting in an
effect reminiscent of a hall of mirrors. We deal with this by including these ``higher order'' multipole moments iteratively until the
difference between successive approximations of $\tilde{U}_{nm}$ drops below a predetermined threshold. In practice, this procedure converged rapidly.

Within the foregoing scheme, the case where one or more electrodes are maintained at a fixed potential other than zero can be treated straightforwardly
by including image charges that contribute to the one-body Hamiltonian rather than to $\tilde{U}_{nm}$. For example, a spherical contact with radius $R$ 
at potential $V$ can
be treated using a hypothetical point charge $q = VR$ at the center of the electrode. This technique is especially useful for
transport calculations in the context of molecular junctions, as it provides the full junction Hamiltonian at finite bias, alleviating the need for
the phenomenological models of capacitive lead-molecule coupling that have been relied upon on in the past.\cite{Stafford09}

\subsection{Screening by a planar electrode}

In classical electrostatics, the image of a charge distribution near a planar conductor is merely the mirror image of the charge
distribution itself. Thus an orbital with monopole moment $q$ and quadrupole moment $Q^{ij}$ located a distance $r$ away from
a conducting plane produces an image orbital inside the conductor located at depth $r$ with multipole moments $\tilde{q} = -q$ and
$\tilde{Q}^{ij} = -\sum_{kl} T_{ik}T_{jl} Q^{kl}$, where $T_{ik}$ is a transformation matrix representing a reflection about a plane parallel
to the surface of the conductor, i.e.
\begin{equation}
\label{reflectionMatrix}
T_{ik} = \delta_{ik} - 2 \hat{n}_i \hat{n}_k,
\end{equation}
where $\hat{n}$ is the unit vector normal to the planar surface.

\subsection{Screening by a spherical electrode}

An orbital with monopole moment $q$ and quadrupole moment $Q^{ij}$ located a distance $r$ from the center of a spherical electrode
with radius $R$ induces an image distribution at $\tilde{r} = R^2/r$ with monopole and quadrupole moments
\[
\tilde{q} =  -q \frac{R}{r} - \frac{R}{2r^3} \sum_{ij} Q^{ij} \hat{r}_i \hat{r}_j
\]
and
\[
\tilde{Q}^{ij} = -\left(\frac{R}{r}\right)^5 \sum_{kl} T_{ik} T_{jl} Q^{kl}
\]
respectively, where $T_{ik}$ is a transformation matrix representing a reflection about the plane normal to the vector $\hat{r}$, similar to Eq.\ \eqref{reflectionMatrix}.

Thus the orbital quadrupole moment induces a higher order image monopole moment, as well as an image of itself that is deformed and reflected.
An image dipole is also generated, but its interaction with the orbital charge distribution is of order $r^{-7}$ and so we have neglected it here.

\section{Screening of the HOMO-LUMO gap}

Although $\pi$-EFT could be used to study a wide variety of phenomena involving conjugated organic molecules, our primary
motivation in formulating it has been to facilitate realistic many-body calculations of transport phenomena in molecular junctions.
In particular, while recent semi-empirical models\cite{Castleton02} reproduce the low-lying excitations of gas-phase benzene, their
predictions of quantities relevant to transport, namely the 
% CAS
fundamental (or HOMO-LUMO) gap and the optical excitations of the ionized molecule, are
less accurate. Moreover, in a molecular junction these quantities are renormalized by screening from metallic electrodes as well as
the presence of linker groups not explicitly included in the molecular Hilbert space. 

Within $\pi$-EFT these effects can be clearly seen: Consider the spectral function of gas-phase benzene, which we evaluate at the many-body level using the non-equilibrium Green's function formalism
 as described in appendix A. Figure \ref{densityOfStatesGasBenzene} shows this quantity, along with experimental values
for the vertical ionization energy ($9.23$ eV), vertical electron affinity ($-1.12$ eV), and the first optical excitation of the cation ($3.04$ eV). As a guide to the eye, the spectrum
has been broadened artificially using a broadening matrix of $\Gamma_{nm} = (0.2 \; \textrm{eV}) \delta_{nm}$. As an aside, we note here that the close agreement between the experimental values and the
maxima of the spectral function suggests our model is accurate at this energy scale. In particular, the accuracy of the theoretical value for the lowest optical excitation of the cation is
noteworthy, as this quantity was not fit during the renormalization procedure but rather represents a prediction of $\pi$-EFT.

\begin{figure}
    \includegraphics[scale=0.95]{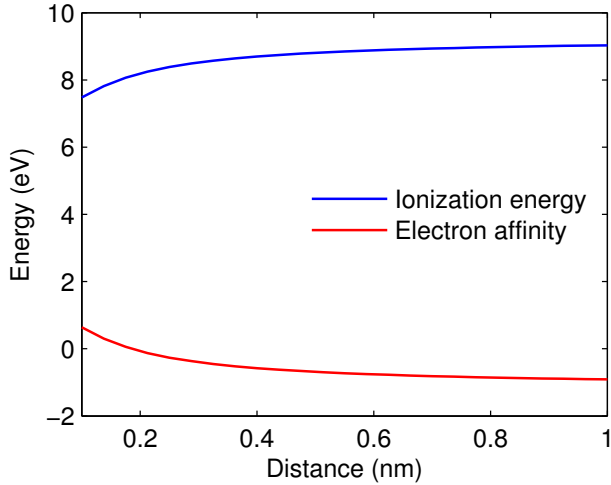}
\caption{\label{screeningOfHOMOLUMOGap}The ionization energy and electron affinity of benzene oriented parallel to the surface of a screening plane, shown as a function of distance.}
\end{figure}

Screening effects become evident when the molecule is brought into proximity with the surface of a planar electrode. Figure \ref{screeningOfHOMOLUMOGap} shows the reduction
of the ionization energy and electron affinity as a function of electrode-molecule distance in this scenario, and the HOMO-LUMO gap, 
given by $IE - EA$, is reduced commensurately. These results, based on the image multipole method, are also consistent with recent GW-based
investigations of screening.\cite{Neaton06,Thygesen09}

\begin{figure}
    \includegraphics[scale=0.95]{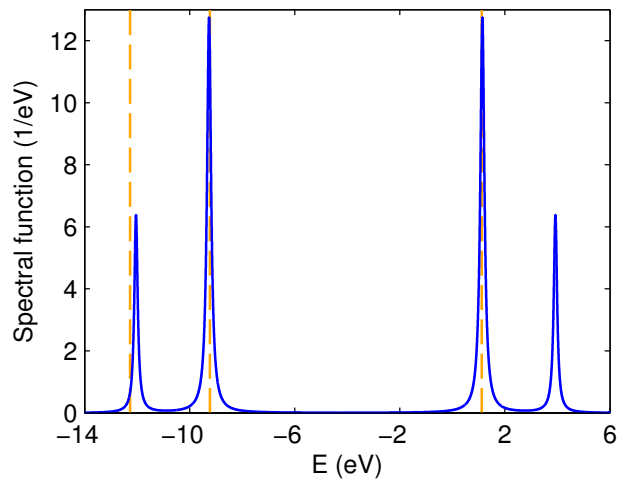}
\caption{\label{densityOfStatesGasBenzene}The spectral function of gas-phase benzene broadened artificially as a guide to the eye. The dashed orange
lines are fixed by (left to right) the lowest lying optical excitation of the molecular cation,\cite{Kovac80, Sell78, Kobayoshi78, Schmidt77, Baltzer97}
the vertical ionization energy of the neutral molecule,\cite{Howell84, Kovac80, Sell78, Kobayoshi78, Schmidt77} and the vertical
electron affinity of the neutral molecule.\cite{Burrow87}}
\end{figure}

\begin{figure}
\vspace{0.5cm}
    \includegraphics[scale=0.75]{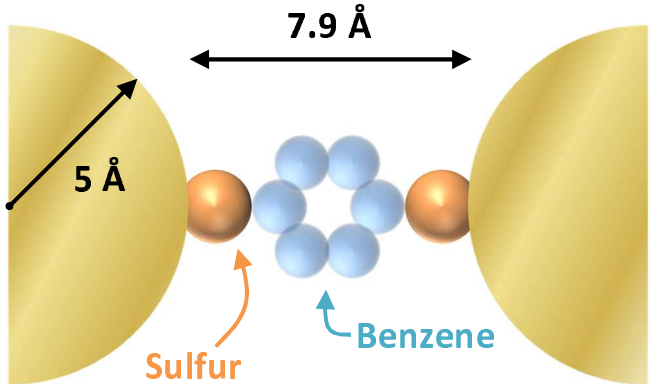}
\caption{\label{geometry}The geometry\cite{Pontes06, letardi04} for the benzenedithiol junction associated with the spectral function shown in Figure \ref{densityOfStatesSpheres}.
The electrodes have been placed so that the screening surface lies one covalent radius\cite{Cordero08} beyond the position of the outermost gold nuclei, a convention that has been investigated
elsewhere\cite{Persson84} in the context of atom-surface van der Waals interactions.}
\end{figure}

We also considered the prototypical benzene-gold junction, consisting
of benzene linked to two gold electrodes via thiol side groups. Although this junction can occur with a wide variety of different geometries, in this
example we have taken the configuration shown in Figure \ref{geometry}. The electrodes are modeled as metallic spheres with radii of $0.5$ nm, and
the partially ionic character of the gold-sulfur bond has been accounted for by placing point charges of $-0.67e$ at the locations of the sulfur atoms. The latter value
was determined in conjunction with the tunneling-width matrix ($\Gamma_{11} = \Gamma_{44} = 0.44$ eV)
via a simultaneous fit of the experimental thermopower\cite{Baheti08} and conductance,\cite{Xiao04} using the techniques described in appendix A. The upper panel of Figure \ref{densityOfStatesSpheres} shows
the spectral function for this junction in the simple case where the tunneling-width matrix is the same as in Figure \ref{densityOfStatesGasBenzene}, a choice which
simplifies comparison of the two cases.

\begin{figure}
    \includegraphics[scale=0.95]{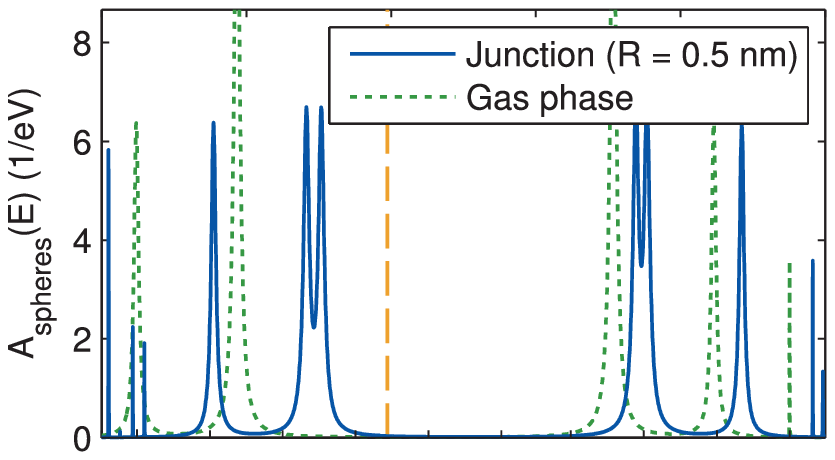}
 \includegraphics[scale=0.95]{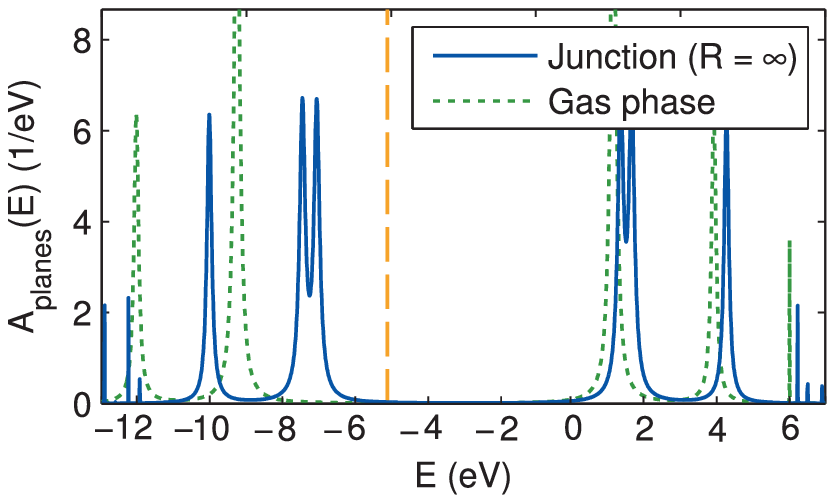}
\caption{\label{densityOfStatesSpheres} Top: The spectral function of the Au-1,4-benzenedithiol-Au junction depicted in Figure \ref{geometry} at room temperature, together with the gas-phase density of states from Figure \ref{densityOfStatesGasBenzene}. To facilitate comparison, the same broadening has been used in both cases. The dotted orange line at $-5.1$ eV indicates the position of the experimental chemical potential of clean gold.\cite{eastman70} Bottom: The spectral function of the same junction with planar instead of spherical  electrodes.}
\end{figure}

Screening from the electrodes reduces the HOMO-LUMO gap by $12.5$ percent as compared to the gas-phase, and the dipole formed by the
gold-sulfur bond shifts the chemical potential of the molecule up by $1.4$ eV. For comparison, we have also calculated the spectral function of the same
junction, but with the electrodes modeled as planes (Figure \ref{densityOfStatesSpheres}, bottom), in which case the screening is maximal and the
HOMO-LUMO gap is reduced by $19$ percent. These results are qualitatively consistent with GW-based investigations of screening effects wherein
a molecule is adsorbed on a metallic surface,\cite{Neaton06, Thygesen09} 
as well as with the recent state-of-the-art GW calculations for benzenedithiol-Au junctions.\cite{Thygesen11}
In comparison to Ref.\ \onlinecite{Thygesen11}, the HOMO and LUMO resonances in Fig.\ \ref{densityOfStatesSpheres} are both shifted slightly upward in energy, but the gap
between them is comparable.  It should be pointed out that the upward shifts of HOMO and LUMO in our model are due in part to the dipole moments of the S-Au bonds, which are
treated phenomenologically in our model, while the screening of the HOMO-LUMO gap is a fundamental effect described by the image multipole method.
%(TO DO: quantitative comparison.)
%though the scenario we have considered is not directly comparable to these works.
As compared to models of screening that treat only the $\pi$-orbital monopole moment,\cite{Kaasbjerg11} the reduction of the HOMO-LUMO gap predicted herein is somewhat smaller, presumably owing to the tendency of the monopole-quadrupole and quadrupole-quadrupole interactions to soften short-range Coulomb interactions. For both of the electrode geometries we considered, a splitting of the two-fold degenerate HOMO and LUMO resonances can also be seen, which arises from the interaction between the $\pi$-electrons and the dipoles associated with the partly ionic gold-sulfur bonds.

We also note that, as compared to DFT-based treatments of similar junctions,\cite{Quek07} the HOMO-LUMO gap seen in Figure \ref{densityOfStatesSpheres} is dramatically larger, consistent with the observation\cite{Quek09} that correlation effects beyond the scope of local DFT must be included to accurately model transport through this junction.

\section{Differential Conductance Spectrum}
\label{sec:dIdV}

\begin{figure*}
    \includegraphics[scale=0.75]{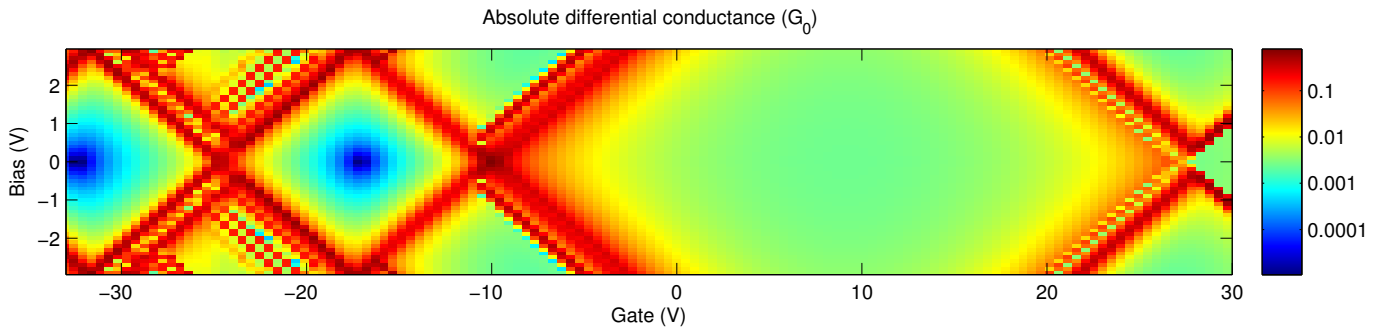}
\caption{\label{conductanceSpheres} Differential conductance spectrum of a Au-1,4-benzenedithiol-Au junction at room temperature versus gate and bias voltages. 
The junction geometry, including source and drain electrodes, is depicted in Fig.\ \ref{geometry}; a spherical gate electrode of radius 3\AA\ (not shown) is centered 5\AA\ 
above the benzene ring.  The effective electrostatic lever arm of the gate is 0.21eV/V.  
The charge on the molecule within the junction is quantized within the diamond-shaped regions centered on the horizontal axis due to the phenomenon
of Coulomb blockade.  Resonant tunneling through electronic excited states at large bias and suppression of transport at small bias due to destructive quantum
interference (blue fringes) are clearly visible.
%Characteristic of Coulomb blockade, the central diamonds correspond to distinct charge states of the molecule.
}
\end{figure*}

The advantages of a computational approach such as $\pi$-EFT combined with many-body NEGF %\cite{Stafford09} 
are perhaps most evident in describing transport through
a molecular junction far from equilibrium.\cite{Kubatkin03,Poot06,Danilov08,Song09}
For then, not only must the equilibrium energetics of electron addition and removal be described correctly, but the dependence of both processes
on both gate and bias voltages must be correct, a significant challenge for conventional approaches.\cite{Datta06}
To illustrate the advantages of $\pi$-EFT %our computational approach 
in this context, we have calculated the differential conductance spectrum of 
a Au-1,4-benzenedithiol-Au junction.  Figure \ref{conductanceSpheres} shows the absolute value of the differential conductance on a logarithmic scale, calculated 
as a function of bias voltage and the electrostatic potential on a spherical gate electrode of radius 3\AA\ centered 5\AA\ above the benzene ring. 
The effective electrostatic lever arm of the gate is 0.21eV/V.
The junction geometry is otherwise identical to that depicted in Fig.\ \ref{geometry}.
In Fig.\ \ref{conductanceSpheres}, we have used the physical tunneling-width matrix $\Gamma_{11} = \Gamma_{44} = 0.44$ eV. 

Of particular note are the diamond-shaped features in the differential conductance spectrum:
the charge on the molecule within the junction is quantized and the differential conductance is suppressed
within the diamond-shaped regions centered along the horizontal axis due to the phenomenon
of Coulomb blockade.\cite{Datta06,Stafford09}  
This is similar to what has been observed experimentally in junctions based 
on larger dithiolated molecules,\cite{Kubatkin03,Poot06,Danilov08} in which case the charging energy is significantly smaller. 
To describe this phenomenon within DFT would require a proper treatment of the derivative discontinuity\cite{Perdew82,Bergfield12} far from equilibrium, for
which no theory currently exists.
To the best of our knowledge, 
charge quantization effects like these are beyond the scope even of self-consistent many-body perturbation theory, e.g.\ as in the 
case of the state-of-the-art DFT + GW approach.\cite{Thygesen11}

Resonant tunneling through electronic excited states at large bias and suppression of transport at small bias due to destructive quantum
interference (blue fringes) are also clearly visible in Fig.\ \ref{conductanceSpheres}.
This differential conductance spectrum %shown in Fig.\ \ref{conductanceSpheres} 
is similar to that obtained previously\cite{Stafford09} 
using a PPP model of the electronic structure.  The main differences are that the sizes of the Coulomb diamonds are reduced due to screening from the metal electrodes, and
the particle-hole symmetry %in Fig.\ \ref{conductanceSpheres} 
of the PPP spectrum is broken by the presence of the S-Au dipoles.

%Of note is the clear presence of a Coulomb blockade diamond, with the HOMO-LUMO gap and
%the HOMO-1 to HOMO spacing roughly equal to $\beta^{-1}(\bar{U} + 2t)$ and $\beta^{-1}(\bar{U})$ respectively, 
%where $\bar{U} = \frac{1}{6^2}\sum_{nm} \tilde{U}_{nm} \approx TODO$ is the screened molecular charging energy and $\beta \approx TODO$ is 
%the approximate electrostatic lever-arm associated with the gate. 

\section{$\pi$-electron contribution to the van der Waals interaction}

As a final application of the image-multipole method,
we consider the $\pi$-electron contribution to the van der Waals interaction between a molecule and a metallic electrode. 
Experimentally, such interactions are important when a molecule is adsorbed on a metal surface, or in single-molecule junctions in
which a molecule bonds directly to metallic electrodes, as in the Pt-benzene-Pt junctions investigated recently by Kiguchi et al.\cite{Kiguchi08} 
Theoretically, 
the van der Waals interaction also represents a unique challenge in that it is a true many-body phenomenon arising
from quantum correlations induced by long-range interactions. As such, it is outside the scope of local approximations to density functional theory,
and modeling van der Waals interactions using nonlocal functionals
is a topic of ongoing research.\cite{Langreth06, Wellendorff10, Michaelides08} 
In contrast to this, the preceding treatment of screening, in conjunction with
a full many-body treatment of the $\pi$-electrons on the molecule, makes it possible to calculate the $\pi$-electron contribution to van der Waals interaction straightforwardly
with no extra adjustable parameters.

In particular, by exactly diagonalizing the few-body molecular 
Hamiltonian with and without the effects of screening included in $U_{nm}$, it is possible to infer the van
der Waals interaction at 
%CAS arbitrary
zero temperature between a molecule and a metallic electrode by comparing the expectation values of
the Hamiltonian in these two cases:
\[
E_{vdW} = \langle \tilde{H} \rangle - \langle H \rangle 
\]

\begin{figure}
    \includegraphics[scale=0.95]{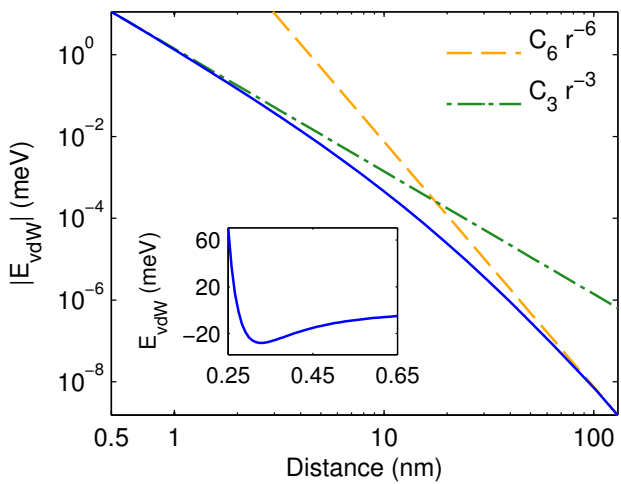}
\caption{\label{vanDerWaalsDistancePlane}The $\pi$-electron contribution to the van der Waals interaction between benzene and a spherical electrode with a radius of $10$ nm, plotted as a function of the
distance from the conducting surface. At all distances the molecule is oriented parallel to the surface of the electrode. The dashed green and orange
lines show the expected asymptotic dependence in the near and far fields respectively. Inset: The same quantity very near
the surface of the electrode, including a phenomenological\cite{Arcidiacono05} gold-carbon hard-core repulsion.}
\end{figure}

\begin{figure}
    \includegraphics[scale=0.95]{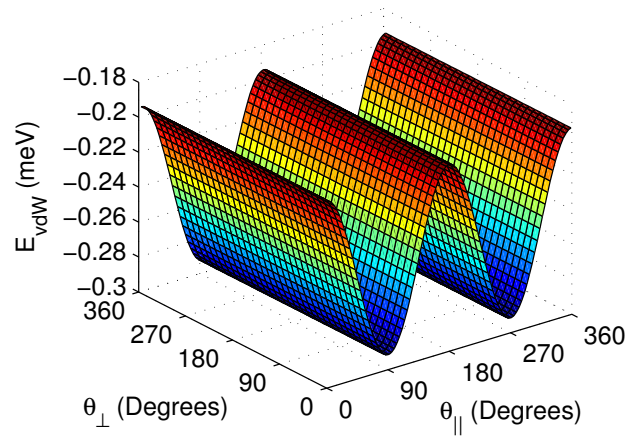}
\caption{\label{vanDerWaalsAnglePlane2nm}The orientation dependence of the $\pi$-electron contribution to van der Waals interaction between a planar electrode and a benzene molecule centered $2$ nm from the metal surface.
The molecule is initially oriented parallel to the electrode and then rotated by an angle $\theta_{\perp}$ about the axis perpendicular to the plane of the molecule,
followed by a rotation of $\theta_{\parallel}$ about an axis within the plane of the molecule.}
\end{figure}

This procedure was carried out at zero temperature for benzene oriented parallel to the surface of a spherical electrode over a large range
of electrode-molecule distances, and the results are shown in Figure \ref{vanDerWaalsDistancePlane}.  When the molecule is near the surface of the electrode $E_{vdW} = -\frac{C_3}{r^3}$, which is the expected asymptotic dependence for the van der Waals interaction
between a molecule and a planar conductor. Conversely, when the molecule is far from the electrode $E_{vdW} = -\frac{C_6}{r^6}$, which is the usual asymptotic dependence
of the van der Waals interaction as given by the Lennard-Jones potential.  A clear transition between the two regimes can be seen around $10$ nm, the radius of the electrode. In the near-field region the constant of proportionality predicted by $\pi$-EFT is $C_3 \approx 1.56$ eV \AA$^3$. 
We also investigated the orientation dependence of the van der Waals interaction between a planar electrode and a benzene molecule, 
as depicted in Figure \ref{vanDerWaalsAnglePlane2nm}, which shows a significantly stronger attractive interaction when the plane of the molecule is
oriented perpendicular to the surface of the electrode.

% MODIFIED: Based on comment 4 from referee 2
The van der Waals coefficient $C_3$ is fundamentally related to the molecular polarizability tensor $\alpha_{ij}$. Thus, for an axially symmetric
molecule such as benzene, a simplified single-oscillator model can be used to derive semi-empirical formulae relating $\alpha_{ij}$ to $C_3$ with the molecule oriented either parallel or perpendicular to the surface of a planar electrode:\cite{Lonij11}
\begin{align}
C^{\parallel}_3 & \approx \frac{E_d}{32}(2\alpha_{\perp} + 2\alpha_{\parallel}) \label{C3parallel} \\
C^{\perp}_3 & \approx \frac{E_d}{32}(\alpha_{\perp} + 3\alpha_{\parallel}) \label{C3perp}
\end{align}
Here $E_d$ is the energy of the principal dipole-allowed optical transition, and $\alpha_\parallel$ and $\alpha_\perp$ are respectively the molecular polarizabilities parallel and perpendicular to its plane of symmetry. As an internal consistency check and to demonstrate that our technique captures the basic physics of the van der Waals interaction, we have calculated these quantities within $\pi$-EFT ($E_d = 7.59$ eV, $\alpha_{\parallel} = 3.24$ \AA$^3$ and $\alpha_{\perp} = 0.00$ \AA$^3$), and used them to deduce $C_3^{\parallel} \approx 1.54$ eV$\cdot$\AA$^3$ and $C^{\perp}_3/C^{\parallel}_3 = 1.5$, which are in close agreement with the values of $C_3$ obtained via direct calculation.

Experimentally, $\alpha_{\parallel} = 12.31$ \AA$^3$, $\alpha_{\perp} = 6.35$ \AA$^3$ and $E_d = 6.93$ eV for benzene, and in this case Eq.\ \eqref{C3parallel} gives $C^{\parallel}_3 \approx 8.08$ eV \AA$^3$, which is roughly five times larger than that predicted by $\pi$-EFT. This discrepancy can be 
attributed to the significant contribution of the $\sigma$-electrons
to the molecular polarizability, as evidenced by the large experimental value of $\alpha_\perp$, which arises from $\sigma$-$\pi$ transitions. Consistent with this and the notion that all of the valence electrons contribute more or less equally to the molecular polarizability, the angular average of the $\pi$-EFT polarizability, i.e. $\frac{\alpha_\perp + 2\alpha_\parallel}{3}$, is roughly a quarter of the same quantity calculated using experimental values. This underscores the importance of the $\sigma$-electron dynamics in the
context of van der Waals interactions, which arise from the long-range spatial correlation of purely virtual processes. In contrast to this, the effect of
the $\sigma$-electrons on real $\pi$-$\pi$ transitions, such as those involved in transport, should be well described by $\pi$-EFT. Moreover, as noted previously, 
the $\sigma$-electrons dynamics can be explicitly included within effective field theory at the expense of a larger
Hilbert space, and we believe that such a $\pi\sigma$-EFT would accurately reproduce the full van der Waals interaction between a conjugated molecule and
a metallic electrode.

\section{Conclusions}

We have shown how EFT can be used to provide a concise derivation of an effective Hamiltonian for $\pi$-electron
systems by performing a multipole expansion, imposing symmetry constraints, and then renormalizing a few adjustable parameters. In particular, we
have optimized the parameters appearing in an effective Hamiltonian for gas-phase benzene, Eq.\ \eqref{benzeneEffectiveHamiltonian}, by 
fitting to experimental data for 1) the vertical ionization energy, 2) the vertical electron affinity, and 3) the six lowest singlet and triplet excitations
of the neutral molecule. This procedure yields a
fit which is comparable to or better than traditional PPP models \cite{Ohno64,Ramasesha91,Castleton02},
and gives $U = 9.69$ eV for the on-site repulsion,
$t = 2.70$ eV for the nearest-neighbor hopping matrix element, $\epsilon = 1.56$ eV for the dielectric constant, and $Q = -0.65$ $e$\AA$^2$ 
for the $\pi$-electron quadrupole moment.  These values of $U$, $t$, and $\epsilon$ are consistent with those used in previous $\pi$-electron models
\cite{Ramasesha91,Castleton02,Wehling11}, while $Q$ is a new physical parameter in our approach, which takes the place
of the ad-hoc functional forms assumed in PPP models and governs the corrections to $1/r$ interactions
at short distances.

We have also utilized $\pi$-EFT to model the 
screening of intramolecular Coulomb interactions by nearby metallic electrodes. 
Within our approach, lead-molecule coupling is treated using a two-step process
wherein all long range Coulomb interactions are included nonperturbatively before lead-molecule tunneling is accounted for
via Dyson's equation. The ability to include finite bias and screening effects via image multipoles--without additional adjustable parameters--represents a significant advantage of
$\pi$-EFT over PPP models, which utilize interactions that do not satisfy Maxwell's equations.

% MODIFIED: Based on comment 5 from referee 2
% MODIFIED: Based on comments 1 and 2 from referee 2
In particular, we have shown how $\pi$-EFT facilitates a realistic description of the prototypical
Au-1,4-benzenedithiol-Au junction, including transport far from equilibrium. 
%In this case, we obtain estimates of both the fundamental (HOMO-LUMO) gap and the chemical potential of the molecule in the junction. 
The accurate description of ionization potential and electron affinity as poles of the Green's function---and their shifts due to interactions with metal electrodes,
sets $\pi$-EFT apart from standard DFT-NEGF approaches, and promises to enable accurate transport calculations for junctions involving 
a variety of conjugated organic molecules.  The ability to simultaneously describe Coulomb blockade and coherent quantum transport appears to set our approach
apart even from state-of-the-art self-consistent many-body perturbation theory.\cite{Thygesen11} The main disadvantages of our approach compared 
to either DFT or DFT + GW are (i) that certain aspects of the junction are described only phenomenologically, such as the linker groups between the molecule and
the metal electrodes; and (ii) that a full diagonalization even of the limited Hilbert space of the $\pi$-electrons scales very poorly.  Nonetheless, exact
diagonalization of $\pi$-EFT should be tractable for conjugated molecules significantly larger than benzene, such as biphenyl or triphenyl, and the use of 
configuration-interaction techniques such as coupled-cluster singles and doubles should allow its application to still larger molecules.
For these systems, 
$\pi$-EFT provides a framework combining an accurate treatment of electron correlation with a higher degree of realism than
is present in conventional PPP techniques.

%Although the exponential growth of the many-body Hilbert space makes it challenging to extend the exact diagonalization techniques of the present 
%work to larger $\pi$-conjugated systems, molecules with at least 18 sites are tractable in principle. In this case, taking into account spin 
%symmetry, a single vector in the largest subspace of the many-body Hilbert space requires on the order of 40 gigabytes of memory to store. 
%Larger systems can still be studied within $\pi$-EFT via approximate treatments of electron correlation, but in this scenario it is unclear 
%if PPP-like methods would retain significant advantages over density functional theory. Nevertheless, nonperturbative interaction effects presently 
%difficult or impossible to treat within DFT are largest in the small molecules to which $\pi$-EFT is most applicable. 

\begin{acknowledgments}
We thank Dr.\ Vincent Lonij for useful discussions.
This material is based upon work supported by the Department of Energy under Award Number DE-SC0006699.
\end{acknowledgments}

\appendix
\section{Many-body theory of transport in molecular junctions}

Within the non-equilibrium Green's function approach to studying transport in molecular junctions, a quantity of central importance is the
retarded Green's function $G$ of the molecule coupled to the electrodes. In the energy domain and using matrix notation, this can be expressed via the Dyson equation as:
\begin{equation}
\label{dyson}
G = G_{mol} + G_{mol} \Sigma G,
\end{equation}
where $G_{mol}$ is the interacting Green's function of the molecule without tunnel coupling to the electrodes, but including long-range Coulomb interactions between the $\pi$-electrons
and their image multipole moments in the leads.
The self-energy $\Sigma$ can be partitioned into the tunneling self-energy $\Sigma_T$ associated with the lead-molecule bonds, and a correction to the Coulomb self-energy
$\Delta \Sigma_C$ arising from lead-molecule coherence:
\[
\Sigma = \Sigma_T + \Delta \Sigma_C.
\]

Far from resonance and at room temperature
$\Delta \Sigma_C \approx 0$, and so in the present context we neglect this correction--an approximation which is justified in detail in ref.\ \citenum{Stafford09}.  
Assuming the leads can be modeled as Fermi liquids with good screening, the electron-electron interactions within them can be neglected and
the tunneling self-energy associated with a given electrode can be expressed as:\cite{DattaBook}
\[
\Sigma_T = Vg(E)V^\dagger,
\]
where $g(E)$ is the retarded Green's function of the lead and $V_{nk}$ are the matrix elements coupling the lead and molecule. In 
the broad-band limit wherein the density of states in the electrodes varies slowly in the vicinity of the metallic Fermi level, the self-energy
then reduces to a purely imaginary matrix with no energy dependence:\cite{DattaBook}
\begin{equation}
\label{tunnelingSelfEnergy}
\Sigma_T = -\frac{i}{2}\sum_{\alpha}\Gamma_{\alpha}.
\end{equation}
Here the tunneling-width matrix $\Gamma_{\alpha}$ associated with lead $\alpha$ given is by
\[
\Gamma_{n\sigma, m\sigma} = 2\pi \rho(\varepsilon_f) V_n V_m^* \delta_{\sigma \sigma'},
\]
where $\rho(\varepsilon_f)$ is the density of states at the metallic Fermi level, and $V_n$ is the matrix element between the $n$th $\pi$-orbital in the molecule and the
lead states in the vicinity of the Fermi level. The diagonal elements of this equation are equivalent to Fermi's golden rule, with $\textrm{Tr} \left\{ \Gamma_{\alpha} /\hbar \right\}$ giving the rate
at which electrons in lead $\alpha$ are being injected into the molecule.

Aside from the self-energy, the other ingredient needed to evaluate Eq.\ \eqref{dyson} is the Green's function of the isolated molecule.
This is determined exactly by first finding the few-body eigenstates $\{|\nu\rangle\}$
and eigenenergies $E_{\nu}$ of the gas-phase molecule, and then using these to explicitly evaluate the molecular Green's function:\cite{Stafford09,Bergfield11}
\begin{equation}
\label{Gmol}
G_{mol} = \sum_{\nu, \nu'} \frac{\left[P(\nu) + P(\nu')\right] C(\nu, \nu')}{E - (E_{\nu'} - E_{\nu}) + i0^+}
\end{equation}
Here $P(\nu)$ is the statistical occupancy of the $\nu$th eigenstate, given at equilibrium by the grand canonical ensemble, and 
\[
C_{n\sigma,m\sigma'}(\nu, \nu') = \langle \nu | d_{n\sigma} | \nu' \rangle \langle \nu' | d^\dagger_{m\sigma'} | \nu \rangle
\]
are many-body matrix elements, where, in the present context, $d^\dagger_{m\sigma}$ creates an electron with spin $\sigma$ in
the $m$th $\pi$-orbital of the molecule.

Altogether, equations \eqref{dyson}, \eqref{tunnelingSelfEnergy} and \eqref{Gmol} provide a method for obtaining the full interacting Green's function of the molecule coupled to the electrodes, which
may then be used to calculate the various physical quantities of interest. For example, the spectral function is given by:\cite{DattaBook}
\[
A(E) = -2 \; \textrm{Im} G,
\]
the trace of which is proportional to the effective single-particle density of states:
\[
\rho(E) = \frac{1}{2\pi} \textrm{Tr} \left\{ A \right\}
\]

Similarly, the elastic transmission function between two electrodes can also be obtained from the full molecular Green's function via the expression:
\[
T_{\alpha \beta} = \textrm{Tr} \left\{ \Gamma_{\alpha} G \Gamma_{\beta} G^\dagger \right\},
\]
where $\Gamma_{\alpha}$ and $\Gamma_{\beta}$ are the tunneling-width matrices associated with leads $\alpha$ and $\beta$
respectively. This quantity may then be used to evaluate the various electronic transport quantities of interest,\cite{Bergfield10} such as the elastic
electrical current
\[
I^{e}_\alpha = \frac{-e}{h}\sum_{\beta} \int dE \; T_{\alpha \beta} \left(f_\beta - f_\alpha\right)
\]
and elastic thermal current
\[
I^Q_\alpha = \frac{1}{h}\sum_{\beta} \int dE \; (E - \mu_{\alpha}) T_{\alpha \beta} \left(f_\beta - f_\alpha\right)
\]
flowing into lead $\alpha$. Here $f_\alpha(E)$ and $\mu_\alpha$ are respectively the Fermi-Dirac distribution and chemical potential
associated with lead $\alpha$.

\bibliographystyle{unsrt}
\bibliography{eftManuscript,refs,DOE_refs}

\end{document}